\documentclass[twocolumn,english]{revtex4-1}
\usepackage[T1]{fontenc}
\usepackage[utf8]{inputenc}
\usepackage{geometry}
\usepackage{float}
\usepackage{graphicx}
\usepackage{xcolor}

\usepackage{babel}
\begin{document}
\title{Are crumpled sheets marginally stable? }

\author{Dor Shohat}
\affiliation{Department of Condensed Matter, School of Physics and Astronomy, Tel Aviv University, Tel Aviv 69978, Israel}
\affiliation{Center for Physics and Chemistry of Living Systems, Tel-Aviv University, Tel Aviv 69978, Israel}

\author{Yoav Lahini}
\affiliation {Department of Condensed Matter, School of Physics and Astronomy, Tel Aviv University, Tel Aviv 69978, Israel}
\affiliation{Center for Physics and Chemistry of Living Systems, Tel-Aviv University, Tel Aviv 69978, Israel}

\author{Daniel Hexner}
\affiliation {Faculty of Mechanical Engineering, Technion, Haifa 32000, Israel}

\begin{abstract}
We study networks of coupled bistable elastic elements, recently proposed as a model for crumpled thin sheets. The networks are poised on the verge of a localized instability, and the model allows unique access to both local and global properties associated with marginal stability. We directly measure pseudo-gaps in the spectrum of local excitations, as well as diverging fluctuations under shear. The networks also host quasi-localized, low-frequency vibrational modes, and scale-free avalanches of instabilities. We propose a correction to the scaling between the pseudo-gap exponent and avalanche statistics based on diverging length fluctuations. Crucially, the dynamics are dominated by a small population of bonds which are locally unstable. Our model combines a coarse-grained view with a continuous, real-space implementation, providing novel insights to a wide class of amorphous solids.

\end{abstract}
\maketitle
\textbf{Introduction: }
The complex dynamics of many disordered systems is governed by localized rearrangements. Systems as diverse as proteins \cite{benichou2011hidden}, amorphous solids \cite{mungan2019networks,keim2021multiperiodic,yan2013glass}, mechanical metamaterials \cite{chen2021reprogrammable,sirote2024emergent}, and thin sheets \cite{bense2021complex,plummer2020buckling,shohat2022memory}, can all be coarse-grained to collections of interacting, multistable-stable elements. This mesoscopic approach has facilitated the understanding of their mechanics and behavior under external manipulation on large scales \cite{nicolas2018deformation}.

In this spirit, it was recently shown that crumpled thin sheets can be mesoscopically modeled as disordered networks of coupled bi-stable springs. This has elucidated the origins of both memory formation under cyclic driving \cite{shohat2022memory}, and intermittent dynamics during creep \cite{shohat2023logarithmic}, behaviors shared by many amorphous solids. Standing alone, each element lies in one of its two minima. However, the incompatibility between rest lengths in the network gives rise to frustration and excess stresses.

Here, we study the energy landscape of this system and reveal unique signatures of marginal behavior. In essence, the system is poised on the verge of undergoing a local instability. As each bond in the network has two stable states, we are motivated to study the model through the prism of spin-glasses \cite{pazmandi1999self}. It has been well appreciated that the energy landscape of glasses is complex and features a pseudo-gap of excitations \cite{mezard1987spin,muller2015marginal} which results in scale-free avalanches of rearrangements under external driving \cite{shang2020elastic,denisov2016universality}.

We show that the model gives exceptional access to local properties that are associated with marginality. It allows for the direct measurement of the pseudo-gap in the excitation spectrum, namely a power-law distribution at small values. The exponents characterizing local strains, stresses, stiffnesses, and energy barriers to instabilities are related via a scaling description \cite{xu2017instabilities}. Marginality results in diverging fluctuations under external shear, and scale-free avalanches. The networks also feature quasi-localized vibrational modes, which act as predictors to instabilities \cite{lerner2021low,richard2020universality}.

Interestingly, we find that there are two populations of bonds, defined by their stability at the single bond level, and characterized by different exponents. A small fraction of bonds ($\sim1\%$) reside in their spinodal region, and have an overwhelming contribution to the spectrum of low-energy excitations. Altogether, we elucidate important features that are often missing or overlooked in disordered mesoscopic models \cite{sollich1997rheology,nicolas2018deformation,kumar2022mapping,sastry2021models}.

\begin{figure}[H]
\begin{centering}
\includegraphics[scale=0.48]{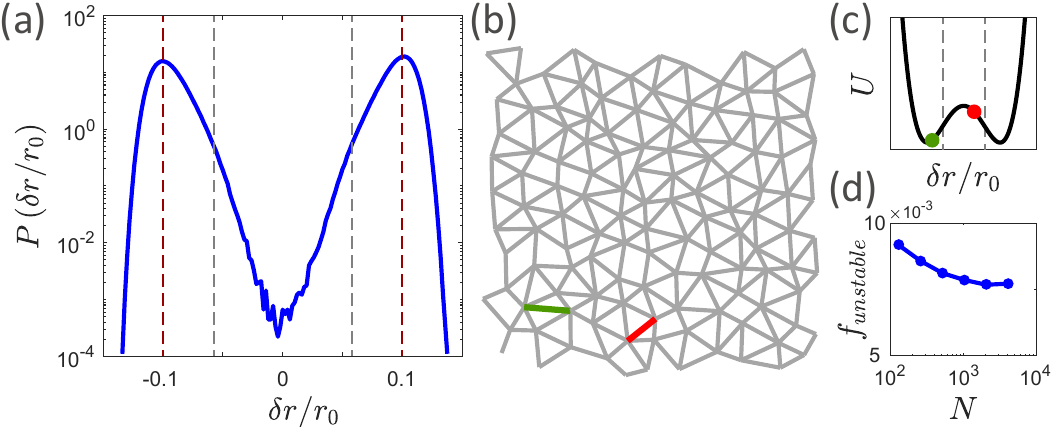}
\par\end{centering}
\vspace*{-0.3cm}
\caption{(a) Bond displacement distribution $P(\delta r/r_{0})$ for $N=4096$. The  red and black dashed lines mark the single bond energy minima and limits of stability respectively. (b) An example of a network. A stable bond is marked in green and an unstable bond is marked in red. Their potential energy is indicated in (c). (d) The fraction of unstable bonds weakly depends on system size. Here, $\alpha/C=0.1$. \label{fig:P_dr}}
\end{figure}

\textbf{Model:} We consider a disordered bonded network with a potential
with two energy minima\citep{shohat2022memory}, corresponding to
a short and long state (see Fig. \ref{fig:P_dr}(b)). The networks are chosen to have a large coordination
number $\Delta Z\equiv Z-4\approx1.5$ to avoid non-generic behaviors
near the rigidity transition, $\Delta Z=0$ \cite{ohern2003jamming}. $Z$ is the coordination number defined
by twice the number of bonds per node. 

The length of the bond is defined by $r=r_{0}+\delta r$ , where $r_{0}$
is the midpoint between the two minima and $\delta r$ is the deviation
from $r_{0}$. For simplicity, we choose the potential to be symmetric around $r_{0}$ (see Fig. \ref{fig:P_dr}(c)):

\begin{equation}
U=\frac{1}{4}C\left(\frac{\delta r}{r_{0}}\right)^{4}-\frac{1}{2}\alpha\left(\frac{\delta r}{r_{0}}\right)^{2}.
\end{equation}
We focus on the case where all bonds have the same parameters, defined
by $C$ and $\alpha$. The locations of the minima can be computed by equating
the force to zero, yielding, $\delta r=\pm\sqrt{\frac{\alpha}{C}}r_{0}.$
The stability of a single bond requires that $U''>0$, which results
in $\left|\delta r/r_{0}\right|>\sqrt{\alpha/3C}$. 

The system is initialized by setting $r_{0}=\ell_{0}\left(1+\delta\right)$,
where $\ell_{0}$ is the initial distance between nodes and $\delta$
is a random number uniformly distributed between $\left[-0.1,0.1\right]$.
We also shear the system back and forth by $10\%$ to prevent possible
non-generic effects of the initial condition. We do not observe a
qualitative dependence on the preparation method. 

\textbf{Characterizing instabilities:} We begin by measuring the distribution
of $\delta r$ shown in Fig. \ref{fig:P_dr}(a). The maximum of the
distribution occurs near the minima $\delta r/r_{0}=\pm\sqrt{a/C}$.
The spread of the distribution results from the internal stresses
that prevent bonds from relaxing to their minima. Since bonds are displaced from their local minima the system is frustrated. 

Interestingly, the distribution of $\delta r$ is non-zero when $\left|\delta r/r_{0}\right|<\sqrt{a/3C}$
where $U''<0$; we refer to these bonds as \emph{unstable}. These
bonds, we will argue, play an important role in determining the stability
of the system. We note that unstable bonds have been previously observed in ordered lattices of bistable springs \citep{nitecki2021mechanical}.
The fraction of unstable bonds as a function of system size is shown
in \ref{fig:P_dr}(d). For systems that are larger than $N\approx500$
nodes the fraction of unstable bonds is approximately constant and
fairly small (less than 1\%). 

While the bonds in the spinodal are unstable at the single bond level
they can be stabilized by the remaining network. Global stability
requires that the Hessian, the matrix of second derivatives, is positive
definite. To gain insight, we consider a simple model where an unstable
bond is attached at both ends with an ordinary spring. The spring
represents a linearization of the elastic contribution of the
remaining network, and its properties are measured. The total energy
as a function of the composite bond is given by the sum of the energy
of the spring with stiffness $k_{sys}$, and the unstable bond with
a potential $U\left(r\right)$: 

\begin{equation}
V\left(r\right)=\frac{k_{sys}}{2}\left(\delta r-a\right)^{2}+U\left(\delta r\right)\label{eq:Composite_bond}
\end{equation}
Force balance on the two springs sets the value of $a$. Stability
depends on the second derivative of the energy, $V''>0$. The composite
bond is stable if $V''\left(r_{0}\right)=k_{sys}+U''\left(r\right)>0$.
An unstable bond with $U''<0$ has a destabilizing effect on the system,
reducing its stiffness, while stable bonds $U''>0$ increase the stiffness
of the system. Thus stable and unstable bonds have a different effect
on the stability of the network. Similar effects have been observed in sticky sphere packings \cite{gonzalez2021mechanical}. 

Based on Eq. \ref{eq:Composite_bond} we can compute the displacement
at which instabilities occur. By explicitly computing the second derivative
of bond $U''$, and setting $V''=0$ we find that the stability boundary
is given by:

\begin{equation}
\left|\frac{\delta r_{in}}{r_{0}}\right|=\sqrt{\frac{\alpha-k_{sys}r_{0}^{2}}{3C}.}\label{eq:disp_insta}
\end{equation}
Thus, the effective stiffness of the network alters the stability
thresholds. Typically, $k_{sys}>0$ and therefore it lowers the thresholds.
If the system is stiff enough $k_{sys}r_{0}^{2}>\alpha$ bistability
is lost, and there is a single minima ($\approx2.5\%$ of bonds). 

We use the approximation presented to compute the displacements to
instabilities. To this
end, we compute for every bond the displacement to reach an instability,
$\Delta r_{in}\equiv\left|\delta r-\delta r_{in}\right|$ using Eq. \ref{eq:disp_insta}. The stiffness $k_{sys}$ is computed from $k_{sys}=V''-U''$, where $V''$ is measured in linear-response.  We note that this is an approximation that relies on linear response, and becomes precise when the distance to the instability vanishes. This approximation neglects non-linear contributions to $k_{sys}$. 

\begin{figure}[H]
\begin{centering}
\vspace*{-0.3cm}

\includegraphics[scale=0.65]{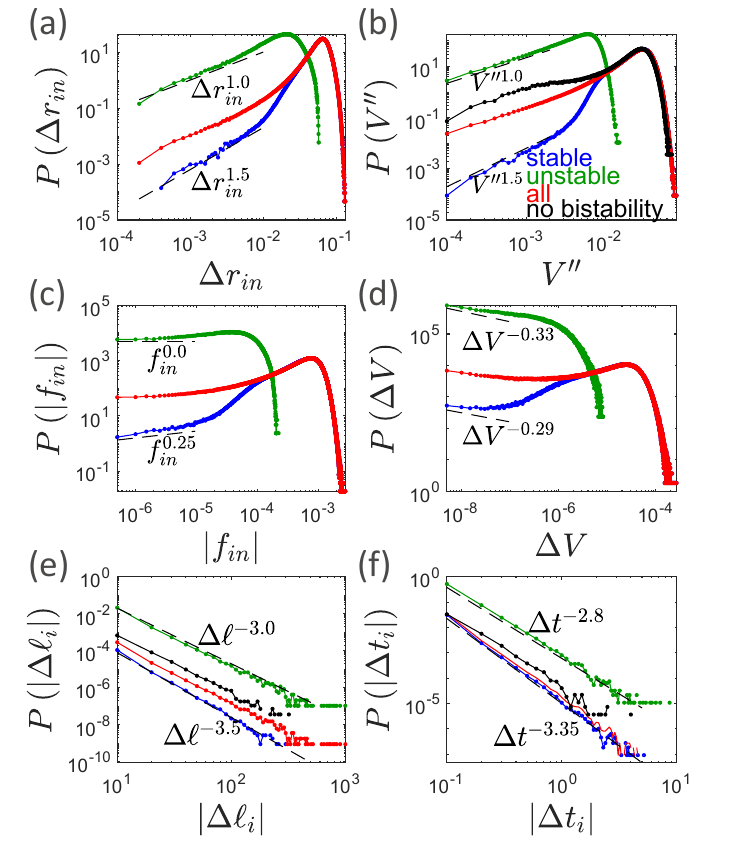}
\par\end{centering}
\vspace*{-0.3cm}
\caption{The distribution of the bond extension to reach an instability (a),  bond stiffness (b), and the force to an instability (c).
(d) The distribution of energy barriers. (e) and (f) are the distributions of susceptibilities, defined as the change in bond lengths (e) and tensions (f) due to shear. Here,
$\alpha/C=0.1$ and $N=512$. \label{fig:Fig2}}
\end{figure}

Fig. \ref{fig:Fig2}(a) shows the distribution of strains to reach
an instability, $\Delta r_{in}.$ At small values the distribution
approximately scales as $\Delta r_{in}^{\approx1.0}$, implying that there
is a quasi-gap. Interestingly, this exponent is consistent with the distribution of local fields to instabilities in the Sherrington-Kirkpatrick model \cite{pazmandi1999self}. We
also consider separately the stable and unstable bonds, indicated
in blue and green. The unstable bonds dominate the distribution despite
being a small fraction of the entire bonds. Surprisingly, the exponent
for the stable bonds appears to be larger $\theta_{stable}\approx1.5$
than that of the unstable bonds $\theta_{unstable}\approx1.0$. This
different behavior for the stable and unstable bonds appears to be
robust. 

Next, we consider the single bond effective stiffness, $V''$, measured
within linear response by squeezing each bond. A vanishing stiffness
is often associated with the nearness to instabilities \citep{xu2010anharmonic}.
Fig. \ref{fig:Fig2}(b) shows the distribution of $V''$ for the stable
(blue), unstable (green) and all the bonds. The distribution at small
values appears to have similar values of exponents measured for $\Delta r_{in}$.
This can be explained by expanding $V''$ in powers of $\Delta r_{in}$,
which yields $V''\propto\Delta r_{in}$. We note, that unlike the
distribution of $\Delta r_{in}$, here, we also include bonds that
are always stable ($k_{sys}r_{0}^{2}>\alpha$). The distribution of
these bonds is shown in black in Fig. \ref{fig:Fig2}(b) and also
have a larger contribution from the stable bonds. In the supporting
information we derive scalings between the different exponents,
consistent with the description of fold instabilities \citep{arnol2003catastrophe,xu2017instabilities}. 

An instability can also be initiated by applying a force that pinches
a bond. We compute the force needed to drive an instability with the condition: 
$f_{in}=V'\left(\delta r_{in}\right).$
Fig. \ref{fig:Fig2}(c) shows that the distribution of $f_{in}$ for
the different populations of bonds. Also here, the unstable and stable
bonds appear to have different exponents. The scaling relations predicts
that $f_{in}\propto\Delta r^{2}$, which predicts that $P\left(f_{in}\right)\propto f_{in}^{\frac{\theta-1}{2}}$.
The unstable bonds at small values scale as $f_{in}^{\approx0}$ while
the stable bonds $f_{in}^{\approx0.25}$, which is consistent with
measurements. Having $P\left(f_{in}\right)\propto f_{in}^{\approx0}$
implies that the system is very susceptible to perturbations. A small
disturbance can lead to an instability. We note however, that these
bonds also have a small stiffness, and as a result local strains may
result only in small changes to the forces. 

Lastly, we compute the distribution of the barriers. We define the barrier
to be the difference between $\Delta V=V\left(\delta r_{in}\right)-V\left(\delta r\right)$.
This is the work needed to initiate an instability. Expanding the
barriers in terms of $\Delta r$ we find that $\Delta V\propto\Delta r^{2+\theta}$.
This predicts that $P\left(\Delta V\right)\propto\Delta V^{\approx-0.33}$
for the unstable bonds and $P\left(\Delta V\right)\propto\Delta V^{\approx-0.29}$
for the stable bonds. 

\textbf{Diverging fluctuations:} The vanishing single bond stiffnesses
suggests that the system is sensitive to small perturbations. Here
we show that the susceptibility is broadly distributed with diverging
moments. We define the local susceptibility by the changes to the bond length or tension, computed within linear-response, due to an applied global shear deformation. 

Fig. \ref{fig:Fig2}(e) shows the distribution of changes in the tensions,
while Fig. \ref{fig:Fig2}(f) shows the distribution of change in
bond lengths. The stable and unstable bonds are considered separately
and appear to have different exponents. The change in tension are
power-law distributed, $P\left(\left|\Delta t_{i}\right|\right)\propto\left|\Delta t_{i}\right|^{\approx-2.8}$
for the unstable bonds and $P\left(\left|\Delta t_{i}\right|\right)\propto\left|\Delta t_{i}\right|^{\approx-3.35}$
for the stable bonds. Also here the unstable bonds dominate the tails
of the distribution. The mean is finite, while the second moment
diverges for the unstable bonds. The distribution of change in bond
length is also power-law, with exponents of $P\left(\Delta\ell_{i}\right)\propto\Delta\ell_{i}^{\approx-3}$
for the unstable bonds and $P\left(\Delta\ell_{i}\right)\propto\Delta\ell_{i}^{\approx-3.5}$
for the stable bonds. Once again, the unstable bonds whose fraction
is very small dominate the distribution. Here, bonds that
have lost their bistability (black) also have a significant contribution. 

A simple scaling can be obtained by assuming that the response is
inversely proportional to the stiffness, $\Delta\ell\propto V''^{-1}$,
which yields $P\left(\Delta\ell\right)\propto\Delta\ell^{-2-\theta}.$
For the unstable bonds, $\theta\approx1$, yielding $\Delta\ell^{-3}$
while for the stable bonds $\theta\approx1.5$ yielding, $\Delta\ell^{-3.5}$.
This agrees with our measurements.

\textbf{Normal mode analysis:} A complementary approach for computing
the low energy excitations relies on the analysis of the eigenmodes
of the system\citep{xu2010anharmonic, manning2011vibrational,kapteijns2020nonlinear,richard2020predicting,franz2015universal}. We
compute the Hessian, the matrix of second derivatives and diagonlize
it. The eigenmodes are the frequencies squared, denoted by $\omega_{i}$.
As discussed, the nearness to an instability is associated with a
vanishing stiffness, and thus small frequencies are predictors
of instabilities. 

\begin{figure}
\begin{centering}
\includegraphics[scale=0.55]{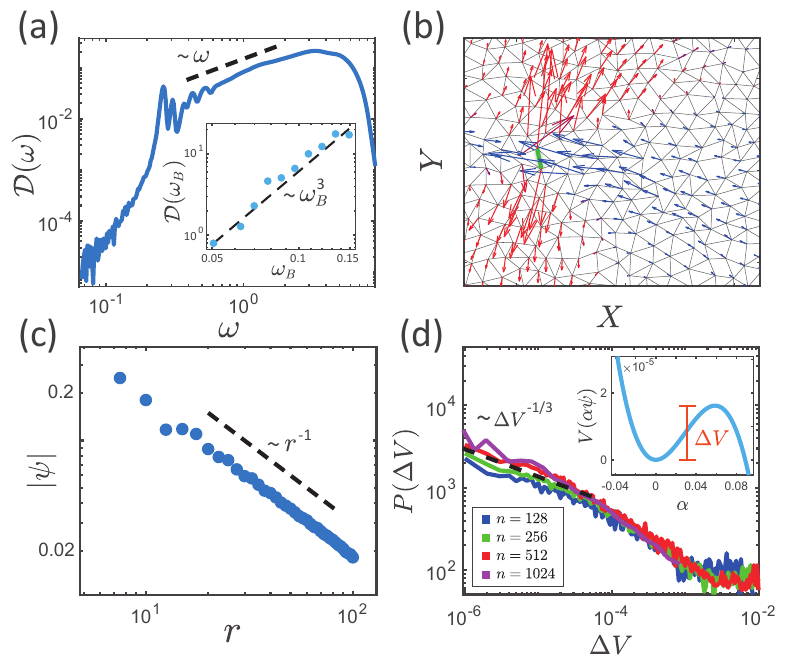}
\par\end{centering}
\vspace*{-0.3cm}
\caption{Linear analysis of instabilities. (a) The density of states features
excess low frequency modes. The density of modes associated with barriers $\omega_B$ scale as $\omega_B^{\approx3.0}$. (b)
An example of a quasi-localized low frequncy mode. It is
centered around an unstable bond. (c) The magnitude of the displacments
decays as $r^{-1}$, consistant with linear elasticity. (d) The distribution
of energy barriers scales approximatly scales as $\Delta V^{-1/3}$
In (a),(b),(c) and the purple curve in (d) $\alpha/C=2.5$. In the
remaining curves in (d) $\alpha/C=0.1$. \label{fig:Linear-analysis}}
\end{figure}

The density of states is shown in Fig. \ref{fig:Linear-analysis}
(a). At intermediate values it is consistent with the Debye scaling $\omega^{1.0}$. At smaller frequencies, we find an excess of quasi-localized modes (QLMs), an example of which is shown in Fig. \ref{fig:Linear-analysis} (b). The
motion is concentrated around a single unstable bond, and the displacement
decays with the distance from that point. To show the localized nature
of these modes, we identify the core for low-frequency normal modes
by the nodes with maximum displacements. We then plot the displacement
field as a function of the distance to the core, shown in Fig. \ref{fig:Linear-analysis}
(c). The displacement field decays as $r^{-1}$, the response to dipole force in linear elasticity.

We now measure the distribution of energy barriers $\Delta V$
by displacing the particles along a normal mode. We consider two approaches
that yield similar behavior: (1) The energy is measured along the
direction of the eigenmode $U(\vec{R}_{eq}+\alpha\psi$), where $R_{eq}$
is the equilibrium state and $\psi$ is the eigenmode vector. The
barrier is identified as the local maxima, illustrated in the inset.
(2) We displace along the eigenmode iteratively and at every step
we minimize the energy to check if the system relaxes to a different
basin of attraction. If the system fails to return to the initial
equilibrium state, we identify the barrier with the difference in
potential energy at the last iteration. Fig. \ref{fig:Linear-analysis}
(d) shows the two methods for different system sizes. At small values
the distribution is consistent with $\Delta V^{-0.33}$.

Finally, we collect only the frequencies of QLMs, $\omega_B$, that are associated with barriers (other QLMs can arise from anomalous local softness \cite{kapteijns2020nonlinear}). Their spectrum scales like  $\omega_B^{\approx3.0}$ (inset of Fig. \ref{fig:Linear-analysis}
(a)). We note that the scaling of $\omega_B^{3.0}$ agrees with the prediction from
the previous analysis where $P\left(V''\right)\propto\left(V''\right)^{1.0}$
at small values. This scaling is at odds with observation of $\omega^{4.0}$
scaling in models of glasses\citep{gurevich2003anharmonicity,richard2020universality,lerner2021low}, presumably due to our preparation protocol \cite{xu2017instabilities}.

\begin{figure}[H]
\begin{centering}
\vspace*{-0.2cm}

\includegraphics[scale=0.55]{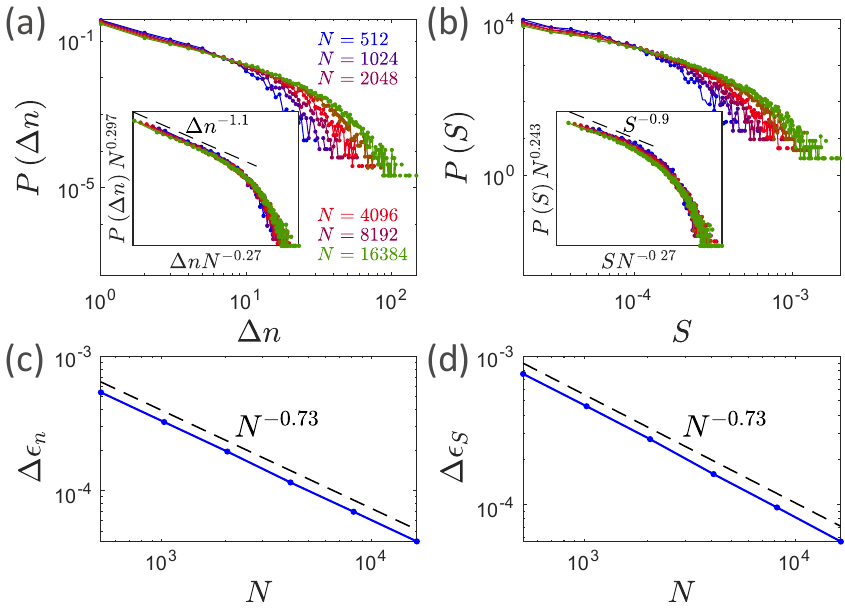}
\par\end{centering}
\vspace*{-0.2cm}

\caption{Characterizing the avalanches based on bond flips (left) and stress drops (right). The top panels show the distributions with a collapse in the inset. The bottom panels show the strain differences between instabilities.
Here, $\alpha/C=0.1$. \label{fig:aval_stress}}
\end{figure}

\textbf{Avalanches:} Next we study the distribution of avalanches,
which are linked to the quasi-gap. Often the quasigap is inferred
from the spacing between avalanches and the avalanche size\citep{karmakar2010statistical,shang2020elastic}.
We consider two independent measurements of the avalanches:\emph{
}1.\emph{ Bond flips:} For each bond we assign a discrete state of
either '+' or '-' depending on the sign of $\delta r$. At each strain
step the number of bonds flips are counted, $\Delta n$. 2.\emph{
Stress drops:} Following Ref. \citep{shang2020elastic} we measure
the size of the avalanche $S=N\left[\Delta u-\Delta\epsilon\sigma\right]$.
Here, $\Delta u$ is the change in energy, $\sigma$ is the stress,
and $\Delta\epsilon$ is the small change in strain. For an elastic
deformation, $S$ is small; large $S$ indicates an instability. We
identify instabilities by setting a cutoff, $S_{c}=10^{-5}$, though
other cutoffs yield similar results. The advantage of this approach
is that it is agnostic to the definition of the instability in terms
of state of the bonds. Fig. \ref{fig:aval_stress}(a) and (b) shows the distribution of the
two types of avalanches. At small values the distribution is approximately
a power-law, decaying approximately as $\Delta n^{\approx-1.1}$ and
$S^{\approx-0.9}$, while at large it features an exponential tail. The size of the avalanches grows with system size. 

To characterize the size of the avalanches we collapse the data by
rescaling the two axis with power of $N$ in the inset of fig. \ref{fig:aval_stress}(a,b) . For both measurements of
avalanches, the average grows as $N^{\approx0.27}$. In addition, we measure the average strain between instabilities $\Delta\epsilon$, shown in
Fig. \ref{fig:aval_stress}(c,d). In both cases, it scales as $N^{\approx-0.73}$.

How does this system size scaling stem from the marginal behaviors described above? We assume that the change in bond length due to a shear deformation scales as $\epsilon \Delta r_{in}^{-\gamma}$. Namely, it is linear in the strain $\epsilon$, but the step size grows when approaching the instability, due to the vanishing stiffness $V''$. An instability occurs if $\Delta r_{in}-b\epsilon \Delta r_{in}^{-\gamma}<0$, where  $b$ is taken to be a constant. Estimating the strain to first instability, using $P\left(\Delta r_{in} \right) \propto \Delta r_{in}^{\theta}$, yields:
\begin{equation}
\Delta \epsilon\propto N^{-\frac{1+\gamma}{1+\theta}}.\label{eq:scaling}
\end{equation}
Note that the standard scaling of $N^{-\frac{1}{1+\theta}}$ \cite{muller2015marginal} is recovered when neglecting the large fluctuations, $\gamma=0$. Similarly, the avalanche size is expected to grow as $N^{\frac{\theta -\gamma}{1+\theta}}$. In the supplementary information, we show that $\gamma\approx0.6$. Altogether, this scaling gives $\theta_{ava}\approx1.19$, close to the direct measurement of $\theta\approx1$. The effect of the two populations of bonds may add further corrections.

\textbf{Conclusions:} We comprehensively studied instabilities in a model of coupled bistable springs. Measurements both at the single bond level, and on a large scale reveal signatures of marginality.
The scaling description of exponents that characterize the proximity to instabilities, where the barrier distribution is $P(\Delta V)\sim (\Delta V^{-1/3})$, resembles the modeling of fold instabilities in amorphous solids \cite{xu2017instabilities}. 

A key observation is a small fraction of bonds in their spinodal regime. These bonds have a destabilizing effect on the system, reducing the stiffness and frequencies. Despite being a small population, the unstable bonds play a pivotal role. They dominate the low-energy excitations, and therefore, we argue, are responsible for initiating avalanches. Moreover, the two bond populations have distinctly different set of exponents; this is reminiscent of excitations in packings \cite{charbonneau2015jamming}.

Importantly, we introduced corrections to the scaling of avalanche sizes $\propto N^{\frac{\theta-\gamma}{1+\theta}}$ and strain intervals between avalanches $\propto N^{-\frac{1+\gamma}{1+\theta}}$. This may affect the measurement of gap exponents in other amorphous systems.

Our work is relevant to understanding the spectrum of excitations in glasses. In particle-based models it is difficult to map out all excitations and measure the nearness to instability \cite{richard2020predicting,lerner2024enumerating}, therefore marginality is often inferred from the distribution of avalanches \cite{shang2020elastic,villarroel2024avalanche}. Here, an exhaustive local study of all instabilities demonstrates new effects that are typically excluded, including distinct populations of excitations and large fluctuations. Perhaps, these effects can be incorporated in future mesoscopic models. Lastly, the large fluctuations in the response could lead to deviations from continuum linear elasticity on large scales. 

Another related problem is memory formation in amorphous solids \cite{paulsen2024mechanical}.
Often it is understood in terms of simplified bistable elements termed \textit{hysterons}, and their interactions \cite{van2021profusion,lindeman2021multiple}. Our work
shows that: 1. The excitation spectrum hosts a non-trivial pseudo-gap which increases the stability of the network. 
Presumably, this facilitates the formation of limit cycles \cite{arceri2021marginal}; 2. Coupling between hysterons alters their stability thresholds, and can even eliminate bistability \cite{szulc2022cooperative}; 3. The spinodal region of individual potentials plays a key role, that may be missed when mapped to a binary variable.

\begin{acknowledgments}
We thank Edan Lerner for insightful discussions. This work was supported by the Israel Science Foundation (grants 2385/20 D.H. and 2117/22 Y.L.) and the Alon Fellowship. D.S. acknowledges support from the Clore Israel Foundation.
\end{acknowledgments}

\bibliography{biblo2}

\begin{thebibliography}{41}%
\makeatletter
\providecommand \@ifxundefined [1]{%
 \@ifx{#1\undefined}
}%
\providecommand \@ifnum [1]{%
 \ifnum #1\expandafter \@firstoftwo
 \else \expandafter \@secondoftwo
 \fi
}%
\providecommand \@ifx [1]{%
 \ifx #1\expandafter \@firstoftwo
 \else \expandafter \@secondoftwo
 \fi
}%
\providecommand \natexlab [1]{#1}%
\providecommand \enquote  [1]{``#1''}%
\providecommand \bibnamefont  [1]{#1}%
\providecommand \bibfnamefont [1]{#1}%
\providecommand \citenamefont [1]{#1}%
\providecommand \href@noop [0]{\@secondoftwo}%
\providecommand \href [0]{\begingroup \@sanitize@url \@href}%
\providecommand \@href[1]{\@@startlink{#1}\@@href}%
\providecommand \@@href[1]{\endgroup#1\@@endlink}%
\providecommand \@sanitize@url [0]{\catcode `\\12\catcode `\$12\catcode `\&12\catcode `\#12\catcode `\^12\catcode `\_12\catcode `\%12\relax}%
\providecommand \@@startlink[1]{}%
\providecommand \@@endlink[0]{}%
\providecommand \url  [0]{\begingroup\@sanitize@url \@url }%
\providecommand \@url [1]{\endgroup\@href {#1}{\urlprefix }}%
\providecommand \urlprefix  [0]{URL }%
\providecommand \Eprint [0]{\href }%
\providecommand \doibase [0]{http://dx.doi.org/}%
\providecommand \selectlanguage [0]{\@gobble}%
\providecommand \bibinfo  [0]{\@secondoftwo}%
\providecommand \bibfield  [0]{\@secondoftwo}%
\providecommand \translation [1]{[#1]}%
\providecommand \BibitemOpen [0]{}%
\providecommand \bibitemStop [0]{}%
\providecommand \bibitemNoStop [0]{.\EOS\space}%
\providecommand \EOS [0]{\spacefactor3000\relax}%
\providecommand \BibitemShut  [1]{\csname bibitem#1\endcsname}%
\let\auto@bib@innerbib\@empty
\bibitem [{\citenamefont {Benichou}\ and\ \citenamefont {Givli}(2011)}]{benichou2011hidden}%
  \BibitemOpen
  \bibfield  {author} {\bibinfo {author} {\bibfnamefont {I.}~\bibnamefont {Benichou}}\ and\ \bibinfo {author} {\bibfnamefont {S.}~\bibnamefont {Givli}},\ }\href@noop {} {\bibfield  {journal} {\bibinfo  {journal} {Applied Physics Letters}\ }\textbf {\bibinfo {volume} {98}} (\bibinfo {year} {2011})}\BibitemShut {NoStop}%
\bibitem [{\citenamefont {Mungan}\ \emph {et~al.}(2019)\citenamefont {Mungan}, \citenamefont {Sastry}, \citenamefont {Dahmen},\ and\ \citenamefont {Regev}}]{mungan2019networks}%
  \BibitemOpen
  \bibfield  {author} {\bibinfo {author} {\bibfnamefont {M.}~\bibnamefont {Mungan}}, \bibinfo {author} {\bibfnamefont {S.}~\bibnamefont {Sastry}}, \bibinfo {author} {\bibfnamefont {K.}~\bibnamefont {Dahmen}}, \ and\ \bibinfo {author} {\bibfnamefont {I.}~\bibnamefont {Regev}},\ }\href@noop {} {\bibfield  {journal} {\bibinfo  {journal} {Physical review letters}\ }\textbf {\bibinfo {volume} {123}},\ \bibinfo {pages} {178002} (\bibinfo {year} {2019})}\BibitemShut {NoStop}%
\bibitem [{\citenamefont {Keim}\ and\ \citenamefont {Paulsen}(2021)}]{keim2021multiperiodic}%
  \BibitemOpen
  \bibfield  {author} {\bibinfo {author} {\bibfnamefont {N.~C.}\ \bibnamefont {Keim}}\ and\ \bibinfo {author} {\bibfnamefont {J.~D.}\ \bibnamefont {Paulsen}},\ }\href@noop {} {\bibfield  {journal} {\bibinfo  {journal} {Science Advances}\ }\textbf {\bibinfo {volume} {7}},\ \bibinfo {pages} {eabg7685} (\bibinfo {year} {2021})}\BibitemShut {NoStop}%
\bibitem [{\citenamefont {Yan}\ \emph {et~al.}(2013)\citenamefont {Yan}, \citenamefont {D{\"u}ring},\ and\ \citenamefont {Wyart}}]{yan2013glass}%
  \BibitemOpen
  \bibfield  {author} {\bibinfo {author} {\bibfnamefont {L.}~\bibnamefont {Yan}}, \bibinfo {author} {\bibfnamefont {G.}~\bibnamefont {D{\"u}ring}}, \ and\ \bibinfo {author} {\bibfnamefont {M.}~\bibnamefont {Wyart}},\ }\href@noop {} {\bibfield  {journal} {\bibinfo  {journal} {Proceedings of the National Academy of Sciences}\ }\textbf {\bibinfo {volume} {110}},\ \bibinfo {pages} {6307} (\bibinfo {year} {2013})}\BibitemShut {NoStop}%
\bibitem [{\citenamefont {Chen}\ \emph {et~al.}(2021)\citenamefont {Chen}, \citenamefont {Pauly},\ and\ \citenamefont {Reis}}]{chen2021reprogrammable}%
  \BibitemOpen
  \bibfield  {author} {\bibinfo {author} {\bibfnamefont {T.}~\bibnamefont {Chen}}, \bibinfo {author} {\bibfnamefont {M.}~\bibnamefont {Pauly}}, \ and\ \bibinfo {author} {\bibfnamefont {P.~M.}\ \bibnamefont {Reis}},\ }\href@noop {} {\bibfield  {journal} {\bibinfo  {journal} {Nature}\ }\textbf {\bibinfo {volume} {589}},\ \bibinfo {pages} {386} (\bibinfo {year} {2021})}\BibitemShut {NoStop}%
\bibitem [{\citenamefont {Sirote-Katz}\ \emph {et~al.}(2024)\citenamefont {Sirote-Katz}, \citenamefont {Shohat}, \citenamefont {Merrigan}, \citenamefont {Lahini}, \citenamefont {Nisoli},\ and\ \citenamefont {Shokef}}]{sirote2024emergent}%
  \BibitemOpen
  \bibfield  {author} {\bibinfo {author} {\bibfnamefont {C.}~\bibnamefont {Sirote-Katz}}, \bibinfo {author} {\bibfnamefont {D.}~\bibnamefont {Shohat}}, \bibinfo {author} {\bibfnamefont {C.}~\bibnamefont {Merrigan}}, \bibinfo {author} {\bibfnamefont {Y.}~\bibnamefont {Lahini}}, \bibinfo {author} {\bibfnamefont {C.}~\bibnamefont {Nisoli}}, \ and\ \bibinfo {author} {\bibfnamefont {Y.}~\bibnamefont {Shokef}},\ }\href@noop {} {\bibfield  {journal} {\bibinfo  {journal} {Nature Communications}\ }\textbf {\bibinfo {volume} {15}},\ \bibinfo {pages} {4008} (\bibinfo {year} {2024})}\BibitemShut {NoStop}%
\bibitem [{\citenamefont {Bense}\ and\ \citenamefont {van Hecke}(2021)}]{bense2021complex}%
  \BibitemOpen
  \bibfield  {author} {\bibinfo {author} {\bibfnamefont {H.}~\bibnamefont {Bense}}\ and\ \bibinfo {author} {\bibfnamefont {M.}~\bibnamefont {van Hecke}},\ }\href@noop {} {\bibfield  {journal} {\bibinfo  {journal} {Proceedings of the National Academy of Sciences}\ }\textbf {\bibinfo {volume} {118}} (\bibinfo {year} {2021})}\BibitemShut {NoStop}%
\bibitem [{\citenamefont {Plummer}\ and\ \citenamefont {Nelson}(2020)}]{plummer2020buckling}%
  \BibitemOpen
  \bibfield  {author} {\bibinfo {author} {\bibfnamefont {A.}~\bibnamefont {Plummer}}\ and\ \bibinfo {author} {\bibfnamefont {D.~R.}\ \bibnamefont {Nelson}},\ }\href@noop {} {\bibfield  {journal} {\bibinfo  {journal} {Physical Review E}\ }\textbf {\bibinfo {volume} {102}},\ \bibinfo {pages} {033002} (\bibinfo {year} {2020})}\BibitemShut {NoStop}%
\bibitem [{\citenamefont {Shohat}\ \emph {et~al.}(2022)\citenamefont {Shohat}, \citenamefont {Hexner},\ and\ \citenamefont {Lahini}}]{shohat2022memory}%
  \BibitemOpen
  \bibfield  {author} {\bibinfo {author} {\bibfnamefont {D.}~\bibnamefont {Shohat}}, \bibinfo {author} {\bibfnamefont {D.}~\bibnamefont {Hexner}}, \ and\ \bibinfo {author} {\bibfnamefont {Y.}~\bibnamefont {Lahini}},\ }\href@noop {} {\bibfield  {journal} {\bibinfo  {journal} {Proceedings of the National Academy of Sciences}\ }\textbf {\bibinfo {volume} {119}},\ \bibinfo {pages} {e2200028119} (\bibinfo {year} {2022})}\BibitemShut {NoStop}%
\bibitem [{\citenamefont {Nicolas}\ \emph {et~al.}(2018)\citenamefont {Nicolas}, \citenamefont {Ferrero}, \citenamefont {Martens},\ and\ \citenamefont {Barrat}}]{nicolas2018deformation}%
  \BibitemOpen
  \bibfield  {author} {\bibinfo {author} {\bibfnamefont {A.}~\bibnamefont {Nicolas}}, \bibinfo {author} {\bibfnamefont {E.~E.}\ \bibnamefont {Ferrero}}, \bibinfo {author} {\bibfnamefont {K.}~\bibnamefont {Martens}}, \ and\ \bibinfo {author} {\bibfnamefont {J.-L.}\ \bibnamefont {Barrat}},\ }\href@noop {} {\bibfield  {journal} {\bibinfo  {journal} {Reviews of Modern Physics}\ }\textbf {\bibinfo {volume} {90}},\ \bibinfo {pages} {045006} (\bibinfo {year} {2018})}\BibitemShut {NoStop}%
\bibitem [{\citenamefont {Shohat}\ \emph {et~al.}(2023)\citenamefont {Shohat}, \citenamefont {Friedman},\ and\ \citenamefont {Lahini}}]{shohat2023logarithmic}%
  \BibitemOpen
  \bibfield  {author} {\bibinfo {author} {\bibfnamefont {D.}~\bibnamefont {Shohat}}, \bibinfo {author} {\bibfnamefont {Y.}~\bibnamefont {Friedman}}, \ and\ \bibinfo {author} {\bibfnamefont {Y.}~\bibnamefont {Lahini}},\ }\href@noop {} {\bibfield  {journal} {\bibinfo  {journal} {Nature Physics}\ }\textbf {\bibinfo {volume} {19}},\ \bibinfo {pages} {1890} (\bibinfo {year} {2023})}\BibitemShut {NoStop}%
\bibitem [{\citenamefont {P{\'a}zm{\'a}ndi}\ \emph {et~al.}(1999)\citenamefont {P{\'a}zm{\'a}ndi}, \citenamefont {Zar{\'a}nd},\ and\ \citenamefont {Zim{\'a}nyi}}]{pazmandi1999self}%
  \BibitemOpen
  \bibfield  {author} {\bibinfo {author} {\bibfnamefont {F.}~\bibnamefont {P{\'a}zm{\'a}ndi}}, \bibinfo {author} {\bibfnamefont {G.}~\bibnamefont {Zar{\'a}nd}}, \ and\ \bibinfo {author} {\bibfnamefont {G.~T.}\ \bibnamefont {Zim{\'a}nyi}},\ }\href@noop {} {\bibfield  {journal} {\bibinfo  {journal} {Physical review letters}\ }\textbf {\bibinfo {volume} {83}},\ \bibinfo {pages} {1034} (\bibinfo {year} {1999})}\BibitemShut {NoStop}%
\bibitem [{\citenamefont {M{\'e}zard}\ \emph {et~al.}(1987)\citenamefont {M{\'e}zard}, \citenamefont {Parisi},\ and\ \citenamefont {Virasoro}}]{mezard1987spin}%
  \BibitemOpen
  \bibfield  {author} {\bibinfo {author} {\bibfnamefont {M.}~\bibnamefont {M{\'e}zard}}, \bibinfo {author} {\bibfnamefont {G.}~\bibnamefont {Parisi}}, \ and\ \bibinfo {author} {\bibfnamefont {M.~A.}\ \bibnamefont {Virasoro}},\ }\href@noop {} {\emph {\bibinfo {title} {Spin glass theory and beyond: An Introduction to the Replica Method and Its Applications}}},\ Vol.~\bibinfo {volume} {9}\ (\bibinfo  {publisher} {World Scientific Publishing Company},\ \bibinfo {year} {1987})\BibitemShut {NoStop}%
\bibitem [{\citenamefont {M{\"u}ller}\ and\ \citenamefont {Wyart}(2015)}]{muller2015marginal}%
  \BibitemOpen
  \bibfield  {author} {\bibinfo {author} {\bibfnamefont {M.}~\bibnamefont {M{\"u}ller}}\ and\ \bibinfo {author} {\bibfnamefont {M.}~\bibnamefont {Wyart}},\ }\href@noop {} {\bibfield  {journal} {\bibinfo  {journal} {Annu. Rev. Condens. Matter Phys.}\ }\textbf {\bibinfo {volume} {6}},\ \bibinfo {pages} {177} (\bibinfo {year} {2015})}\BibitemShut {NoStop}%
\bibitem [{\citenamefont {Shang}\ \emph {et~al.}(2020)\citenamefont {Shang}, \citenamefont {Guan},\ and\ \citenamefont {Barrat}}]{shang2020elastic}%
  \BibitemOpen
  \bibfield  {author} {\bibinfo {author} {\bibfnamefont {B.}~\bibnamefont {Shang}}, \bibinfo {author} {\bibfnamefont {P.}~\bibnamefont {Guan}}, \ and\ \bibinfo {author} {\bibfnamefont {J.-L.}\ \bibnamefont {Barrat}},\ }\href@noop {} {\bibfield  {journal} {\bibinfo  {journal} {Proceedings of the National Academy of Sciences}\ }\textbf {\bibinfo {volume} {117}},\ \bibinfo {pages} {86} (\bibinfo {year} {2020})}\BibitemShut {NoStop}%
\bibitem [{\citenamefont {Denisov}\ \emph {et~al.}(2016)\citenamefont {Denisov}, \citenamefont {L{\"o}rincz}, \citenamefont {Uhl}, \citenamefont {Dahmen},\ and\ \citenamefont {Schall}}]{denisov2016universality}%
  \BibitemOpen
  \bibfield  {author} {\bibinfo {author} {\bibfnamefont {D.}~\bibnamefont {Denisov}}, \bibinfo {author} {\bibfnamefont {K.}~\bibnamefont {L{\"o}rincz}}, \bibinfo {author} {\bibfnamefont {J.}~\bibnamefont {Uhl}}, \bibinfo {author} {\bibfnamefont {K.~A.}\ \bibnamefont {Dahmen}}, \ and\ \bibinfo {author} {\bibfnamefont {P.}~\bibnamefont {Schall}},\ }\href@noop {} {\bibfield  {journal} {\bibinfo  {journal} {Nature communications}\ }\textbf {\bibinfo {volume} {7}},\ \bibinfo {pages} {10641} (\bibinfo {year} {2016})}\BibitemShut {NoStop}%
\bibitem [{\citenamefont {Xu}\ \emph {et~al.}(2017)\citenamefont {Xu}, \citenamefont {Liu},\ and\ \citenamefont {Nagel}}]{xu2017instabilities}%
  \BibitemOpen
  \bibfield  {author} {\bibinfo {author} {\bibfnamefont {N.}~\bibnamefont {Xu}}, \bibinfo {author} {\bibfnamefont {A.~J.}\ \bibnamefont {Liu}}, \ and\ \bibinfo {author} {\bibfnamefont {S.~R.}\ \bibnamefont {Nagel}},\ }\href@noop {} {\bibfield  {journal} {\bibinfo  {journal} {Physical review letters}\ }\textbf {\bibinfo {volume} {119}},\ \bibinfo {pages} {215502} (\bibinfo {year} {2017})}\BibitemShut {NoStop}%
\bibitem [{\citenamefont {Lerner}\ and\ \citenamefont {Bouchbinder}(2021)}]{lerner2021low}%
  \BibitemOpen
  \bibfield  {author} {\bibinfo {author} {\bibfnamefont {E.}~\bibnamefont {Lerner}}\ and\ \bibinfo {author} {\bibfnamefont {E.}~\bibnamefont {Bouchbinder}},\ }\href@noop {} {\bibfield  {journal} {\bibinfo  {journal} {The Journal of chemical physics}\ }\textbf {\bibinfo {volume} {155}} (\bibinfo {year} {2021})}\BibitemShut {NoStop}%
\bibitem [{\citenamefont {Richard}\ \emph {et~al.}(2020{\natexlab{a}})\citenamefont {Richard}, \citenamefont {Gonz{\'a}lez-L{\'o}pez}, \citenamefont {Kapteijns}, \citenamefont {Pater}, \citenamefont {Vaknin}, \citenamefont {Bouchbinder},\ and\ \citenamefont {Lerner}}]{richard2020universality}%
  \BibitemOpen
  \bibfield  {author} {\bibinfo {author} {\bibfnamefont {D.}~\bibnamefont {Richard}}, \bibinfo {author} {\bibfnamefont {K.}~\bibnamefont {Gonz{\'a}lez-L{\'o}pez}}, \bibinfo {author} {\bibfnamefont {G.}~\bibnamefont {Kapteijns}}, \bibinfo {author} {\bibfnamefont {R.}~\bibnamefont {Pater}}, \bibinfo {author} {\bibfnamefont {T.}~\bibnamefont {Vaknin}}, \bibinfo {author} {\bibfnamefont {E.}~\bibnamefont {Bouchbinder}}, \ and\ \bibinfo {author} {\bibfnamefont {E.}~\bibnamefont {Lerner}},\ }\href@noop {} {\bibfield  {journal} {\bibinfo  {journal} {Physical Review Letters}\ }\textbf {\bibinfo {volume} {125}},\ \bibinfo {pages} {085502} (\bibinfo {year} {2020}{\natexlab{a}})}\BibitemShut {NoStop}%
\bibitem [{\citenamefont {Sollich}\ \emph {et~al.}(1997)\citenamefont {Sollich}, \citenamefont {Lequeux}, \citenamefont {H{\'e}braud},\ and\ \citenamefont {Cates}}]{sollich1997rheology}%
  \BibitemOpen
  \bibfield  {author} {\bibinfo {author} {\bibfnamefont {P.}~\bibnamefont {Sollich}}, \bibinfo {author} {\bibfnamefont {F.}~\bibnamefont {Lequeux}}, \bibinfo {author} {\bibfnamefont {P.}~\bibnamefont {H{\'e}braud}}, \ and\ \bibinfo {author} {\bibfnamefont {M.~E.}\ \bibnamefont {Cates}},\ }\href@noop {} {\bibfield  {journal} {\bibinfo  {journal} {Physical review letters}\ }\textbf {\bibinfo {volume} {78}},\ \bibinfo {pages} {2020} (\bibinfo {year} {1997})}\BibitemShut {NoStop}%
\bibitem [{\citenamefont {Kumar}\ \emph {et~al.}(2022)\citenamefont {Kumar}, \citenamefont {Patinet}, \citenamefont {Maloney}, \citenamefont {Regev}, \citenamefont {Vandembroucq},\ and\ \citenamefont {Mungan}}]{kumar2022mapping}%
  \BibitemOpen
  \bibfield  {author} {\bibinfo {author} {\bibfnamefont {D.}~\bibnamefont {Kumar}}, \bibinfo {author} {\bibfnamefont {S.}~\bibnamefont {Patinet}}, \bibinfo {author} {\bibfnamefont {C.~E.}\ \bibnamefont {Maloney}}, \bibinfo {author} {\bibfnamefont {I.}~\bibnamefont {Regev}}, \bibinfo {author} {\bibfnamefont {D.}~\bibnamefont {Vandembroucq}}, \ and\ \bibinfo {author} {\bibfnamefont {M.}~\bibnamefont {Mungan}},\ }\href@noop {} {\bibfield  {journal} {\bibinfo  {journal} {The Journal of Chemical Physics}\ }\textbf {\bibinfo {volume} {157}} (\bibinfo {year} {2022})}\BibitemShut {NoStop}%
\bibitem [{\citenamefont {Sastry}(2021)}]{sastry2021models}%
  \BibitemOpen
  \bibfield  {author} {\bibinfo {author} {\bibfnamefont {S.}~\bibnamefont {Sastry}},\ }\href@noop {} {\bibfield  {journal} {\bibinfo  {journal} {Physical Review Letters}\ }\textbf {\bibinfo {volume} {126}},\ \bibinfo {pages} {255501} (\bibinfo {year} {2021})}\BibitemShut {NoStop}%
\bibitem [{\citenamefont {O’hern}\ \emph {et~al.}(2003)\citenamefont {O’hern}, \citenamefont {Silbert}, \citenamefont {Liu},\ and\ \citenamefont {Nagel}}]{ohern2003jamming}%
  \BibitemOpen
  \bibfield  {author} {\bibinfo {author} {\bibfnamefont {C.~S.}\ \bibnamefont {O’hern}}, \bibinfo {author} {\bibfnamefont {L.~E.}\ \bibnamefont {Silbert}}, \bibinfo {author} {\bibfnamefont {A.~J.}\ \bibnamefont {Liu}}, \ and\ \bibinfo {author} {\bibfnamefont {S.~R.}\ \bibnamefont {Nagel}},\ }\href@noop {} {\bibfield  {journal} {\bibinfo  {journal} {Physical Review E}\ }\textbf {\bibinfo {volume} {68}},\ \bibinfo {pages} {011306} (\bibinfo {year} {2003})}\BibitemShut {NoStop}%
\bibitem [{\citenamefont {Nitecki}\ and\ \citenamefont {Givli}(2021)}]{nitecki2021mechanical}%
  \BibitemOpen
  \bibfield  {author} {\bibinfo {author} {\bibfnamefont {S.}~\bibnamefont {Nitecki}}\ and\ \bibinfo {author} {\bibfnamefont {S.}~\bibnamefont {Givli}},\ }\href@noop {} {\bibfield  {journal} {\bibinfo  {journal} {Journal of the Mechanics and Physics of Solids}\ }\textbf {\bibinfo {volume} {157}},\ \bibinfo {pages} {104634} (\bibinfo {year} {2021})}\BibitemShut {NoStop}%
\bibitem [{\citenamefont {Gonz{\'a}lez-L{\'o}pez}\ \emph {et~al.}(2021)\citenamefont {Gonz{\'a}lez-L{\'o}pez}, \citenamefont {Shivam}, \citenamefont {Zheng}, \citenamefont {Ciamarra},\ and\ \citenamefont {Lerner}}]{gonzalez2021mechanical}%
  \BibitemOpen
  \bibfield  {author} {\bibinfo {author} {\bibfnamefont {K.}~\bibnamefont {Gonz{\'a}lez-L{\'o}pez}}, \bibinfo {author} {\bibfnamefont {M.}~\bibnamefont {Shivam}}, \bibinfo {author} {\bibfnamefont {Y.}~\bibnamefont {Zheng}}, \bibinfo {author} {\bibfnamefont {M.~P.}\ \bibnamefont {Ciamarra}}, \ and\ \bibinfo {author} {\bibfnamefont {E.}~\bibnamefont {Lerner}},\ }\href@noop {} {\bibfield  {journal} {\bibinfo  {journal} {Physical Review E}\ }\textbf {\bibinfo {volume} {103}},\ \bibinfo {pages} {022605} (\bibinfo {year} {2021})}\BibitemShut {NoStop}%
\bibitem [{\citenamefont {Xu}\ \emph {et~al.}(2010)\citenamefont {Xu}, \citenamefont {Vitelli}, \citenamefont {Liu},\ and\ \citenamefont {Nagel}}]{xu2010anharmonic}%
  \BibitemOpen
  \bibfield  {author} {\bibinfo {author} {\bibfnamefont {N.}~\bibnamefont {Xu}}, \bibinfo {author} {\bibfnamefont {V.}~\bibnamefont {Vitelli}}, \bibinfo {author} {\bibfnamefont {A.~J.}\ \bibnamefont {Liu}}, \ and\ \bibinfo {author} {\bibfnamefont {S.~R.}\ \bibnamefont {Nagel}},\ }\href@noop {} {\bibfield  {journal} {\bibinfo  {journal} {EPL (Europhysics Letters)}\ }\textbf {\bibinfo {volume} {90}},\ \bibinfo {pages} {56001} (\bibinfo {year} {2010})}\BibitemShut {NoStop}%
\bibitem [{\citenamefont {Arnol'd}(2003)}]{arnol2003catastrophe}%
  \BibitemOpen
  \bibfield  {author} {\bibinfo {author} {\bibfnamefont {V.~I.}\ \bibnamefont {Arnol'd}},\ }\href@noop {} {\emph {\bibinfo {title} {Catastrophe theory}}}\ (\bibinfo  {publisher} {Springer Science \& Business Media},\ \bibinfo {year} {2003})\BibitemShut {NoStop}%
\bibitem [{\citenamefont {Manning}\ and\ \citenamefont {Liu}(2011)}]{manning2011vibrational}%
  \BibitemOpen
  \bibfield  {author} {\bibinfo {author} {\bibfnamefont {M.~L.}\ \bibnamefont {Manning}}\ and\ \bibinfo {author} {\bibfnamefont {A.~J.}\ \bibnamefont {Liu}},\ }\href@noop {} {\bibfield  {journal} {\bibinfo  {journal} {Physical Review Letters}\ }\textbf {\bibinfo {volume} {107}},\ \bibinfo {pages} {108302} (\bibinfo {year} {2011})}\BibitemShut {NoStop}%
\bibitem [{\citenamefont {Kapteijns}\ \emph {et~al.}(2020)\citenamefont {Kapteijns}, \citenamefont {Richard},\ and\ \citenamefont {Lerner}}]{kapteijns2020nonlinear}%
  \BibitemOpen
  \bibfield  {author} {\bibinfo {author} {\bibfnamefont {G.}~\bibnamefont {Kapteijns}}, \bibinfo {author} {\bibfnamefont {D.}~\bibnamefont {Richard}}, \ and\ \bibinfo {author} {\bibfnamefont {E.}~\bibnamefont {Lerner}},\ }\href@noop {} {\bibfield  {journal} {\bibinfo  {journal} {Physical Review E}\ }\textbf {\bibinfo {volume} {101}},\ \bibinfo {pages} {032130} (\bibinfo {year} {2020})}\BibitemShut {NoStop}%
\bibitem [{\citenamefont {Richard}\ \emph {et~al.}(2020{\natexlab{b}})\citenamefont {Richard}, \citenamefont {Ozawa}, \citenamefont {Patinet}, \citenamefont {Stanifer}, \citenamefont {Shang}, \citenamefont {Ridout}, \citenamefont {Xu}, \citenamefont {Zhang}, \citenamefont {Morse}, \citenamefont {Barrat}, \citenamefont {Berthier}, \citenamefont {Falk}, \citenamefont {Guan}, \citenamefont {Liu}, \citenamefont {Martens}, \citenamefont {Sastry}, \citenamefont {Vandembroucq}, \citenamefont {Lerner},\ and\ \citenamefont {Manning}}]{richard2020predicting}%
  \BibitemOpen
  \bibfield  {author} {\bibinfo {author} {\bibfnamefont {D.}~\bibnamefont {Richard}}, \bibinfo {author} {\bibfnamefont {M.}~\bibnamefont {Ozawa}}, \bibinfo {author} {\bibfnamefont {S.}~\bibnamefont {Patinet}}, \bibinfo {author} {\bibfnamefont {E.}~\bibnamefont {Stanifer}}, \bibinfo {author} {\bibfnamefont {B.}~\bibnamefont {Shang}}, \bibinfo {author} {\bibfnamefont {S.~A.}\ \bibnamefont {Ridout}}, \bibinfo {author} {\bibfnamefont {B.}~\bibnamefont {Xu}}, \bibinfo {author} {\bibfnamefont {G.}~\bibnamefont {Zhang}}, \bibinfo {author} {\bibfnamefont {P.~K.}\ \bibnamefont {Morse}}, \bibinfo {author} {\bibfnamefont {J.-L.}\ \bibnamefont {Barrat}}, \bibinfo {author} {\bibfnamefont {L.}~\bibnamefont {Berthier}}, \bibinfo {author} {\bibfnamefont {M.~L.}\ \bibnamefont {Falk}}, \bibinfo {author} {\bibfnamefont {P.}~\bibnamefont {Guan}}, \bibinfo {author} {\bibfnamefont {A.~J.}\ \bibnamefont {Liu}}, \bibinfo {author} {\bibfnamefont {K.}~\bibnamefont {Martens}}, \bibinfo {author} {\bibfnamefont {S.}~\bibnamefont
  {Sastry}}, \bibinfo {author} {\bibfnamefont {D.}~\bibnamefont {Vandembroucq}}, \bibinfo {author} {\bibfnamefont {E.}~\bibnamefont {Lerner}}, \ and\ \bibinfo {author} {\bibfnamefont {M.~L.}\ \bibnamefont {Manning}},\ }\href@noop {} {\bibfield  {journal} {\bibinfo  {journal} {Physical Review Materials}\ }\textbf {\bibinfo {volume} {4}},\ \bibinfo {pages} {113609} (\bibinfo {year} {2020}{\natexlab{b}})}\BibitemShut {NoStop}%
\bibitem [{\citenamefont {Franz}\ \emph {et~al.}(2015)\citenamefont {Franz}, \citenamefont {Parisi}, \citenamefont {Urbani},\ and\ \citenamefont {Zamponi}}]{franz2015universal}%
  \BibitemOpen
  \bibfield  {author} {\bibinfo {author} {\bibfnamefont {S.}~\bibnamefont {Franz}}, \bibinfo {author} {\bibfnamefont {G.}~\bibnamefont {Parisi}}, \bibinfo {author} {\bibfnamefont {P.}~\bibnamefont {Urbani}}, \ and\ \bibinfo {author} {\bibfnamefont {F.}~\bibnamefont {Zamponi}},\ }\href@noop {} {\bibfield  {journal} {\bibinfo  {journal} {Proceedings of the National Academy of Sciences}\ }\textbf {\bibinfo {volume} {112}},\ \bibinfo {pages} {14539} (\bibinfo {year} {2015})}\BibitemShut {NoStop}%
\bibitem [{\citenamefont {Gurevich}\ \emph {et~al.}(2003)\citenamefont {Gurevich}, \citenamefont {Parshin},\ and\ \citenamefont {Schober}}]{gurevich2003anharmonicity}%
  \BibitemOpen
  \bibfield  {author} {\bibinfo {author} {\bibfnamefont {V.}~\bibnamefont {Gurevich}}, \bibinfo {author} {\bibfnamefont {D.}~\bibnamefont {Parshin}}, \ and\ \bibinfo {author} {\bibfnamefont {H.}~\bibnamefont {Schober}},\ }\href@noop {} {\bibfield  {journal} {\bibinfo  {journal} {Physical Review B}\ }\textbf {\bibinfo {volume} {67}},\ \bibinfo {pages} {094203} (\bibinfo {year} {2003})}\BibitemShut {NoStop}%
\bibitem [{\citenamefont {Karmakar}\ \emph {et~al.}(2010)\citenamefont {Karmakar}, \citenamefont {Lerner},\ and\ \citenamefont {Procaccia}}]{karmakar2010statistical}%
  \BibitemOpen
  \bibfield  {author} {\bibinfo {author} {\bibfnamefont {S.}~\bibnamefont {Karmakar}}, \bibinfo {author} {\bibfnamefont {E.}~\bibnamefont {Lerner}}, \ and\ \bibinfo {author} {\bibfnamefont {I.}~\bibnamefont {Procaccia}},\ }\href@noop {} {\bibfield  {journal} {\bibinfo  {journal} {Physical Review E}\ }\textbf {\bibinfo {volume} {82}},\ \bibinfo {pages} {055103} (\bibinfo {year} {2010})}\BibitemShut {NoStop}%
\bibitem [{\citenamefont {Charbonneau}\ \emph {et~al.}(2015)\citenamefont {Charbonneau}, \citenamefont {Corwin}, \citenamefont {Parisi},\ and\ \citenamefont {Zamponi}}]{charbonneau2015jamming}%
  \BibitemOpen
  \bibfield  {author} {\bibinfo {author} {\bibfnamefont {P.}~\bibnamefont {Charbonneau}}, \bibinfo {author} {\bibfnamefont {E.~I.}\ \bibnamefont {Corwin}}, \bibinfo {author} {\bibfnamefont {G.}~\bibnamefont {Parisi}}, \ and\ \bibinfo {author} {\bibfnamefont {F.}~\bibnamefont {Zamponi}},\ }\href@noop {} {\bibfield  {journal} {\bibinfo  {journal} {Physical review letters}\ }\textbf {\bibinfo {volume} {114}},\ \bibinfo {pages} {125504} (\bibinfo {year} {2015})}\BibitemShut {NoStop}%
\bibitem [{\citenamefont {Lerner}\ \emph {et~al.}(2024)\citenamefont {Lerner}, \citenamefont {Moriel},\ and\ \citenamefont {Bouchbinder}}]{lerner2024enumerating}%
  \BibitemOpen
  \bibfield  {author} {\bibinfo {author} {\bibfnamefont {E.}~\bibnamefont {Lerner}}, \bibinfo {author} {\bibfnamefont {A.}~\bibnamefont {Moriel}}, \ and\ \bibinfo {author} {\bibfnamefont {E.}~\bibnamefont {Bouchbinder}},\ }\href@noop {} {\bibfield  {journal} {\bibinfo  {journal} {arXiv preprint arXiv:2404.12735}\ } (\bibinfo {year} {2024})}\BibitemShut {NoStop}%
\bibitem [{\citenamefont {Villarroel}\ and\ \citenamefont {D{\"u}ring}(2024)}]{villarroel2024avalanche}%
  \BibitemOpen
  \bibfield  {author} {\bibinfo {author} {\bibfnamefont {C.}~\bibnamefont {Villarroel}}\ and\ \bibinfo {author} {\bibfnamefont {G.}~\bibnamefont {D{\"u}ring}},\ }\href@noop {} {\bibfield  {journal} {\bibinfo  {journal} {Soft Matter}\ }\textbf {\bibinfo {volume} {20}},\ \bibinfo {pages} {3520} (\bibinfo {year} {2024})}\BibitemShut {NoStop}%
\bibitem [{\citenamefont {Paulsen}\ and\ \citenamefont {Keim}(2024)}]{paulsen2024mechanical}%
  \BibitemOpen
  \bibfield  {author} {\bibinfo {author} {\bibfnamefont {J.~D.}\ \bibnamefont {Paulsen}}\ and\ \bibinfo {author} {\bibfnamefont {N.~C.}\ \bibnamefont {Keim}},\ }\href@noop {} {\bibfield  {journal} {\bibinfo  {journal} {arXiv preprint arXiv:2405.08158}\ } (\bibinfo {year} {2024})}\BibitemShut {NoStop}%
\bibitem [{\citenamefont {van Hecke}(2021)}]{van2021profusion}%
  \BibitemOpen
  \bibfield  {author} {\bibinfo {author} {\bibfnamefont {M.}~\bibnamefont {van Hecke}},\ }\href@noop {} {\bibfield  {journal} {\bibinfo  {journal} {Physical Review E}\ }\textbf {\bibinfo {volume} {104}},\ \bibinfo {pages} {054608} (\bibinfo {year} {2021})}\BibitemShut {NoStop}%
\bibitem [{\citenamefont {Lindeman}\ and\ \citenamefont {Nagel}(2021)}]{lindeman2021multiple}%
  \BibitemOpen
  \bibfield  {author} {\bibinfo {author} {\bibfnamefont {C.~W.}\ \bibnamefont {Lindeman}}\ and\ \bibinfo {author} {\bibfnamefont {S.~R.}\ \bibnamefont {Nagel}},\ }\href@noop {} {\bibfield  {journal} {\bibinfo  {journal} {Science Advances}\ }\textbf {\bibinfo {volume} {7}},\ \bibinfo {pages} {eabg7133} (\bibinfo {year} {2021})}\BibitemShut {NoStop}%
\bibitem [{\citenamefont {Arceri}\ \emph {et~al.}(2021)\citenamefont {Arceri}, \citenamefont {Corwin},\ and\ \citenamefont {Hagh}}]{arceri2021marginal}%
  \BibitemOpen
  \bibfield  {author} {\bibinfo {author} {\bibfnamefont {F.}~\bibnamefont {Arceri}}, \bibinfo {author} {\bibfnamefont {E.~I.}\ \bibnamefont {Corwin}}, \ and\ \bibinfo {author} {\bibfnamefont {V.~F.}\ \bibnamefont {Hagh}},\ }\href@noop {} {\bibfield  {journal} {\bibinfo  {journal} {Physical Review E}\ }\textbf {\bibinfo {volume} {104}},\ \bibinfo {pages} {044907} (\bibinfo {year} {2021})}\BibitemShut {NoStop}%
\bibitem [{\citenamefont {Szulc}\ \emph {et~al.}(2022)\citenamefont {Szulc}, \citenamefont {Mungan},\ and\ \citenamefont {Regev}}]{szulc2022cooperative}%
  \BibitemOpen
  \bibfield  {author} {\bibinfo {author} {\bibfnamefont {A.}~\bibnamefont {Szulc}}, \bibinfo {author} {\bibfnamefont {M.}~\bibnamefont {Mungan}}, \ and\ \bibinfo {author} {\bibfnamefont {I.}~\bibnamefont {Regev}},\ }\href@noop {} {\bibfield  {journal} {\bibinfo  {journal} {The Journal of Chemical Physics}\ }\textbf {\bibinfo {volume} {156}} (\bibinfo {year} {2022})}\BibitemShut {NoStop}%
\end{thebibliography}%

\end{document}